\documentclass[prb,twocolumn,superscriptaddress]{revtex4-1}
\usepackage{amsmath}
\usepackage{amssymb}
\usepackage{amsthm}
\usepackage{amsfonts}
\usepackage{listings}
\usepackage{enumerate}
\usepackage{latexsym}
\usepackage{color}
\usepackage{bm}
\usepackage{hyperref}
\hypersetup{
 pdfnewwindow=true, colorlinks=true,
 linkcolor=blue, anchorcolor=blue,
 citecolor=blue, filecolor=blue,
 menucolor=blue, urlcolor=blue}

\newcommand{\beginsupplement}{%
        \setcounter{table}{0}
        \renewcommand{\thetable}{S\arabic{table}}%
        \setcounter{figure}{0}
        \renewcommand{\thefigure}{S\arabic{figure}}%
     }

\usepackage{psfrag}

\usepackage{bm}
\usepackage{graphicx}
\usepackage{subfigure}

\newcommand{\webirvsp}{\href{https://github.com/zjwang11/irvsp/blob/master/src_irvsp_v2.tar.gz}{\ttfamily IRVSP}}

\RequirePackage[normalem]{ulem} 
\RequirePackage{color}\definecolor{RED}{rgb}{1,0,0}\definecolor{BLUE}{rgb}{0,0,1} 

\newcommand{\ket}[1]{\left|#1\right\rangle}

\def\ie{{\it i.e.},\ }

\input{epsf}

\begin{document}

\tolerance 10000

\newcommand{\vk}{{\bf k}}

\draft

\title{Large spin Hall conductivity and excellent hydrogen evolution reaction activity in unconventional PtTe$_{1.75}$ monolayer}

\author{Dexi Shao}
\affiliation{Beijing National Laboratory for Condensed Matter Physics,
and Institute of Physics, Chinese Academy of Sciences, Beijing 100190, China}
\affiliation{Department of Physics, Hangzhou Normal University, Hangzhou 311121, China}

\author{Junze Deng}
\affiliation{Beijing National Laboratory for Condensed Matter Physics,
and Institute of Physics, Chinese Academy of Sciences, Beijing 100190, China}
\affiliation{University of Chinese Academy of Sciences, Beijing 100049, China}

\author{Haohao Sheng}
\affiliation{Beijing National Laboratory for Condensed Matter Physics,
and Institute of Physics, Chinese Academy of Sciences, Beijing 100190, China}
\affiliation{University of Chinese Academy of Sciences, Beijing 100049, China}

\author{Ruihan Zhang}
\affiliation{Beijing National Laboratory for Condensed Matter Physics,
and Institute of Physics, Chinese Academy of Sciences, Beijing 100190, China}
\affiliation{University of Chinese Academy of Sciences, Beijing 100049, China}

\author{Hongming Weng}
\affiliation{Beijing National Laboratory for Condensed Matter Physics,
and Institute of Physics, Chinese Academy of Sciences, Beijing 100190, China}
\affiliation{University of Chinese Academy of Sciences, Beijing 100049, China}

\author{Zhong Fang}
\affiliation{Beijing National Laboratory for Condensed Matter Physics,
and Institute of Physics, Chinese Academy of Sciences, Beijing 100190, China}
\affiliation{University of Chinese Academy of Sciences, Beijing 100049, China}

\author{Xing-Qiu Chen}

\author{Yan Sun}
\email{sunyan@imr.ac.cn}
\affiliation{Shenyang National Laboratory for Materials Science, Institute of Metal Research,
 Chinese Academy of Science, 110016 Shenyang, Liaoning, People's Republic of China}
\affiliation{School of Materials Science and Engineering, University of Science and Technology of China}

\author{Zhijun Wang}
\email{wzj@iphy.ac.cn}
\affiliation{Beijing National Laboratory for Condensed Matter Physics,
and Institute of Physics, Chinese Academy of Sciences, Beijing 100190, China}
\affiliation{University of Chinese Academy of Sciences, Beijing 100049, China}

\begin{abstract}
Two-dimensional (2D) materials have gained lots of attention due to the potential applications. In this work, we propose that based on first-principles calculations, the (2$\times$2) patterned PtTe$_2$ monolayer with kagome lattice formed by the well-ordered Te vacancy (PtTe$_{1.75}$) hosts large spin Hall conductivity (SHC) and excellent hydrogen evolution reaction (HER) activity.
The unconventional nature relies on the $A1@1b$ band representation (BR) of the highest valence band without SOC.
The large SHC comes from the Rashba spin-orbit coupling (SOC) in the noncentrosymmetric structure induced by the Te vacancy.
Even though it has a metallic SOC band structure, the $\mathbb Z_2$ invariant is well defined due to the existence of the direct band gap and is computed to be nontrivial.
The calculated SHC is as large as 1.25$\times 10^3 \frac{\hbar}{e} (\Omega~cm)^{-1}$ at the Fermi level ($E_F$).
By tuning the chemical potential from $E_F-0.3$ to $E_F+0.3$ eV, it varies rapidly and monotonically from $-1.2\times 10^3$ to 3.1$\times 10^3 \frac{\hbar}{e} (\Omega~cm)^{-1}$.
In addition, we also find the Te vacancy in the patterned monolayer can induce excellent HER activity.
Our results not only offer a new idea to search 2D materials with large SHC, i.e., by introducing inversion-symmetry breaking vacancies in large SOC systems, but also provide a feasible system with tunable SHC (by applying gate voltage) and excellent HER activity.
\end{abstract}

\maketitle
\section{Introduction}
In the past decade, many topological semimetals with various quasi-particle dispersions and fascinating properties have been proposed~\cite{Ashvin2018RMP,Wan2011,DSM-Cd3As2,Bradlyn2016Beyond,Wieder2016DDSM}.
The layered noble transition-metal dichalcogenide PtTe$_2$ is extraordinary with heavily tilted type-II Dirac fermion~\cite{Yan2017}.
It hosts unique properties, such as topological nontrivial band structure~\cite{Yan2017,3DPlasmons2018}, ultrahigh electrical conductivity~\cite{Fu2018,Hao2018}, and robustness of the remaining semimetal phase even down to just two triatomic layers~\cite{PhysRevLett.124.036402,DENG20191044}.
Soon after, many PtTe$_2$ derivatives have been proposed, including the monolayer, multilayer, doping, vacancy, heterojunction structures and so on. For example, the Ir-doped PtTe$_2$ (\ie Pt$_{1-x}$Ir$_{x}$Te$_2$) has realized the Fermi level ($E_F$) tunability and superconductivity, which opens up a new route for the investigation of Dirac physics and topological superconductivity~\cite{SC2018,Jiang2020,PhysRevB.98.125143}. More recently, PtTe$_2$-based broadband photodetectors and image sensors have been fabricated, demonstrating tremendous potential application value in various photoelectric devices~\cite{Tong2020,Shawkat2021,PtTe2vdwHeter2019}. Very recently, the patterned monolayer with kagome lattice formed by one Te vacancy in the $2\times 2$ supercell has been grown successfully~\cite{PtTe2-epitaxial-2021}, whose band topology and potential properties are unknown. The study of PtTe$_2$ derivatives can not only reveal novel condensed matter physics but also facilitate the versatile development in device physics.

\begin{figure}[!tb]
\centering
\includegraphics[width=8.5 cm]{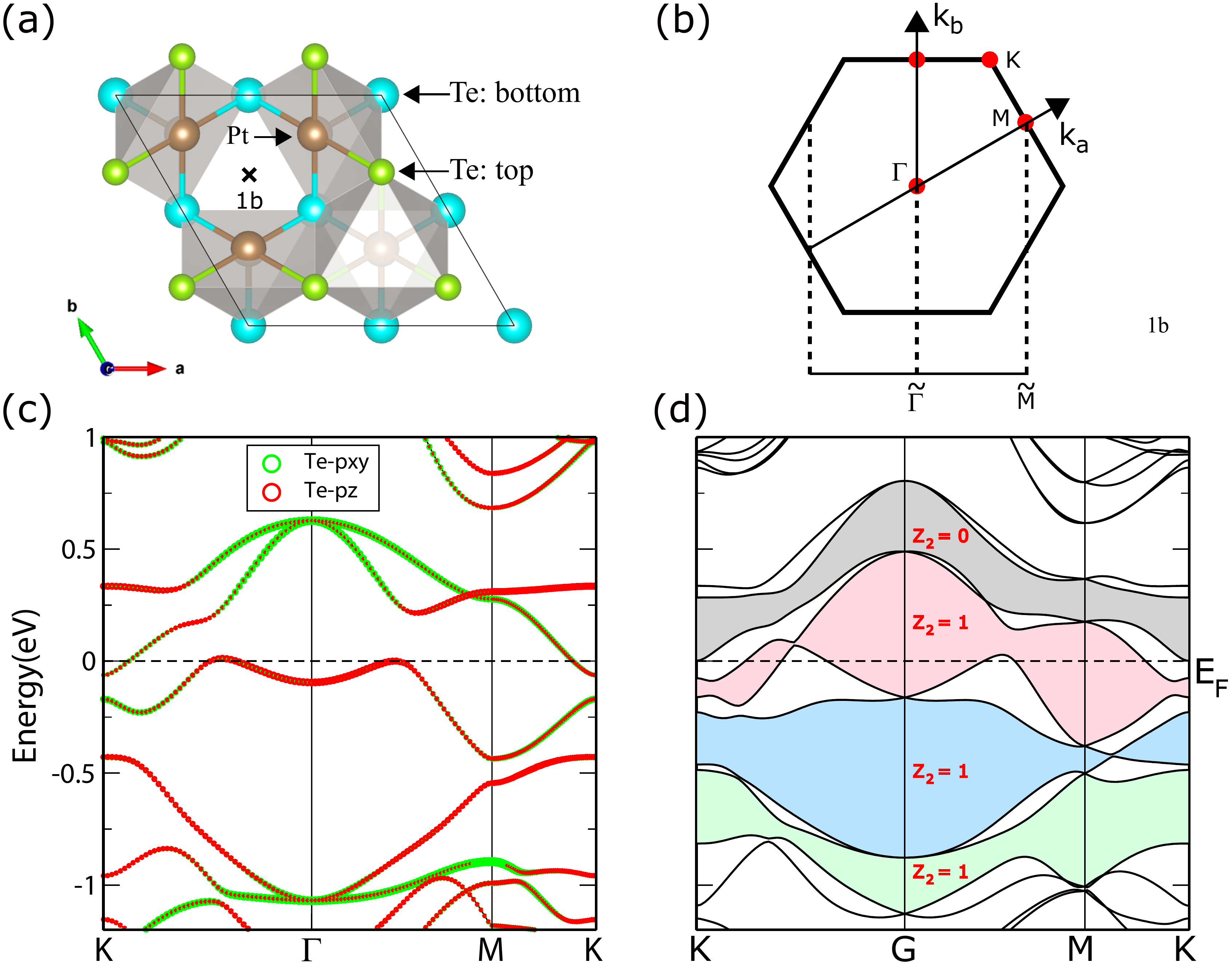}
\caption{(Color online)
(a)Crystal structure of the patterned PtTe$_2$ monolayer (PtTe$_{1.75}$).
The pristine PtTe$_2$ monolayer system in the kagome lattice contains two Te layers, with 4 Te atoms in both the bottom layer and the top layer of the (2$\times$2) supercell.
And the PtTe$_{1.75}$ system comes from the patterned PtTe$_2$ monolayer with a well-ordered Te vacancy (schematized by ``$\times$'' at $1b$ Wyckoff site) in the top layer of the (2$\times$2) supercell.
Thus, there are 3 Te atoms (denoted by green ball) occupying in the top layer while 4 Te atoms (denoted by blue ball) occupying in the bottom layer in the PtTe$_{1.75}$ system.
(b) The corresponding 2D bulk BZ and one-dimensional (1D) projected BZ orthogonal to the (01) edge.
Band structures of the PtTe$_{1.75}$ system (c) without and (d) with SOC. The light green, blue, red and grey zones in Fig.~\ref{fig:1}(d) indicate that there exist direct band gaps between the corresponding adjacent bands.
Thus, the time reversal $\mathbb Z_2$ can be defined and calculated to be 1, 1, 1 and 0 with $(N_e-4)$, $(N_e-2)$, $N_e$ and $(N_e+2)$ occupied bands.
} \label{fig:1}
\end{figure}

\begin{figure*}[!tb]
\centering
\includegraphics[width=12 cm]{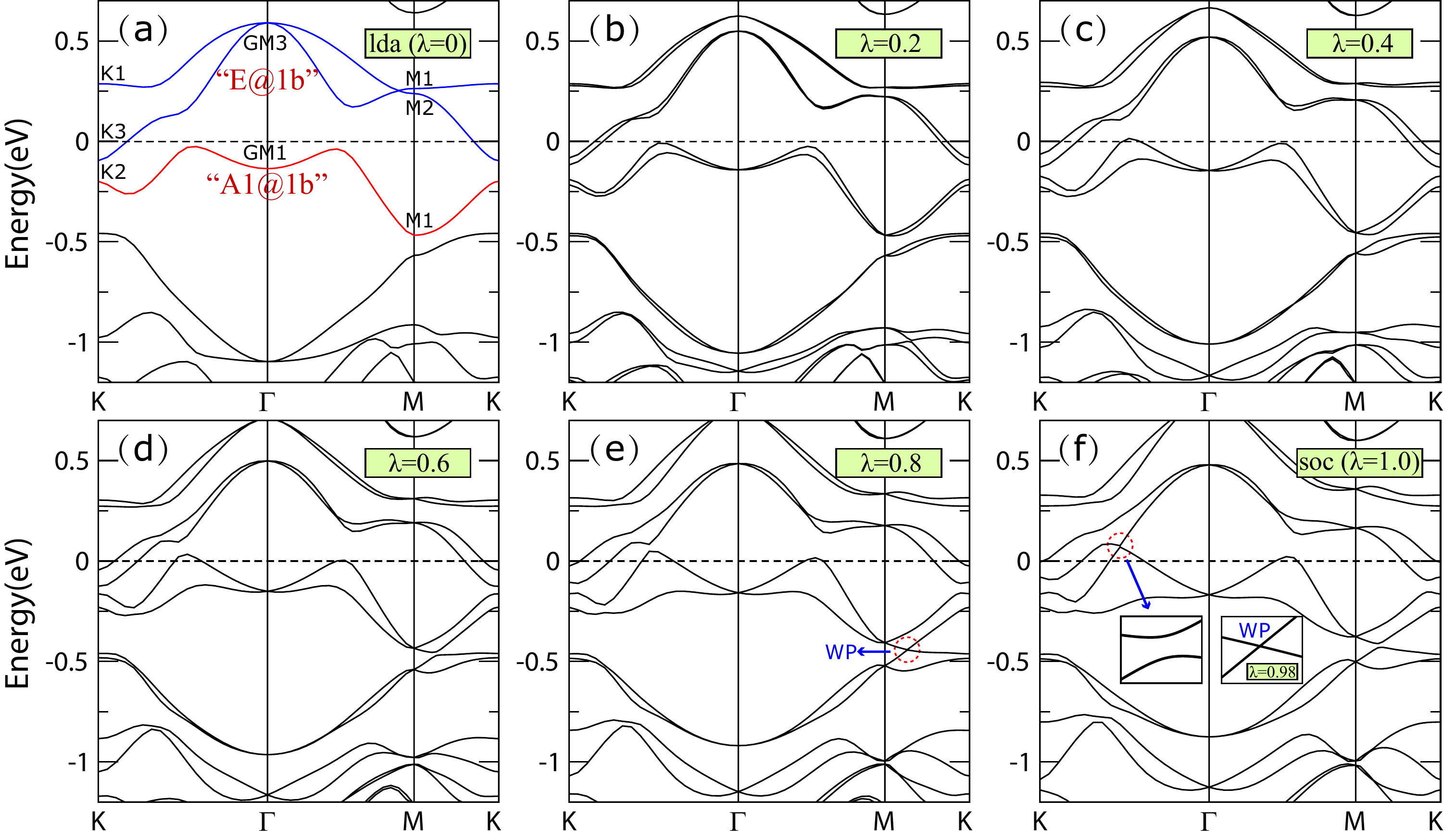}
\caption{(Color online)
Band structures of PtTe$_{1.75}$ monolayer with the strength of SOC (a) $\lambda = 0$ (without SOC), (b) $\lambda = 0.2$, (c) $\lambda = 0.4$, (d) $\lambda = 0.6$, (e) $\lambda = 0.8$ and (f) $\lambda = 1.0$ (with actual SOC).
For the case without SOC shown in Fig.~\ref{fig:2}(a), the two conduction bands schematized by two blue lines belong to $E@1b$ BR, while the highest valence band schematized by the red line belong to $A1@1b$ BR, respectively.
There exists a WP along the M--K line below $E_F$ when $\lambda = 0.8$. Bands near E$_F$ undergoes a gap closing and reopening progress when the strength of SOC evolves from 0.0 to 1.0, which gives a topological nontrivial band gap with SOC ($\lambda = 1.0$). The critical transition occurs at $\lambda = 0.98$ (right inset in the $\lambda = 1.0$ panel), which gives another WP along the K-$\Gamma$ line.
} \label{fig:2}
\end{figure*}


In this work, we theoretically propose that the recently synthesized patterned PtTe$_2$ monolayer with the Te vacancy hosts large SHC due to the Rashba SOC, where the Te vacancy breaks inversion symmetry ($\mathcal{I}$).
The momentum offset  and strength of the Rashba SOC are estimated, $k_0=0.12$ \AA$^{-1}$ and $\alpha_{R}=0.8112$ eV {\AA}.
The momentum offset  $k_0$ is very large and comparable with the largest one reported in the Bi/Ag(111) surface alloy in literature~\cite{Ag111-Bi-2007}, which induces visible Rashba band splitting.
Using the Kubo formula approach at the clean limit, we find the Rashba SOC will induce large SHC, as large as 1.25$\times 10^3 \frac{\hbar}{e} (\Omega~cm)^{-1}$ at $E_F$. Furthermore, the SHC changes rapidly and monotonically as the chemical potential evolving in a wide range ($E-E_F\in[-0.3,0.3]$ eV), which benefits to the potential applications in spintronics. In the end, the variation of the Gibbs free energy for hydrogen adsorption progress is considered, which indicates that the Te vacancy can induce a superior HER activity.

\section{Crystal structure and Methodology}
The pristine PtTe$_2$ crystallizes in the CdI$_2$-type trigonal ($1T$) structure with $P\bar{3}m1$ space group (SG). It hosts the layered structure stacking along the $z$ axis. The patterned monolayer with kagome lattice formed by one Te vacancy in the 2$\times$2 supercell has been successfully grown on the Pt(111) surface~\cite{PtTe2-epitaxial-2021}.
As shown in Fig.~\ref{fig:1}(a), the patterned PtTe$_2$ monolayer (\ie PtTe$_{1.75}$) contains two Te layers: four Te atoms (blue balls) in the bottom layer and three Te atoms (green balls; with one vacancy schematized by ``$\times$'' at $1b$ Wyckoff site) in the top layer. The distance between the bottom and top layers is $d_0=2.7253$~\AA.
The Te vacancy breaks $\mathcal{I}$, resulting in a noncentrosymmetrical structure with p3m1 layer group (LG 69; corresponding to SG $P3m1$ excluding translational symmetry along the $z$ axis). Thus, the Rashba SOC induced band splitting is inevitable.
The lattice parameters and atomic positions are listed in Table~\ref{table:Xnet} [in Section A of the Supplemental Material (SM ~\ref{supone})].

We performed the first-principles calculations based on the density functional theory (DFT) using projector augmented wave (PAW) method \cite{paw1,paw2} implemented in the Vienna \emph{ab initio} simulation package (VASP)~\cite{KRESSE199615,vasp}.
The generalized gradient approximation (GGA) with exchange-correlation functional of Perdew, Burke and Ernzerhof (PBE) for  the exchange-correlation functional \cite{pbe} was adopted.
The kinetic energy cutoff  was set to 500 eV for the plane wave bases. The thickness of the vacuum layer along $z$ axis was set to $>20$ \AA.
The BZ was sampled by $\Gamma$-centered Monkhorst-Pack method with a 12$\times$12$\times$1 $\textbf{k}$-mesh for the 2D periodic boundary conditions in the self-consistent process. The Wilson-loop technique~\cite{Yu2011An} was used to calculate the $\mathbb Z_2$ topological invariant. 
In addition, the electronic structures near $E_F$ were doubly checked by the full-potential local-orbital code (FPLO)~\cite{FPLO1999} and fully consistent with that from VASP. The wannier-based tight-binding (TB) model under bases of the Te-$p$ and Pt-$d$ orbitals is extracted from the DFT calculations for the calculations of SHC.

\section{Calculations and Results}

\subsection{Electronic band structures}
The band structures without and with SOC are presented in Fig.~\ref{fig:1}(c) and Fig.~\ref{fig:1}(d), respectively.
Comparing them, we notice that the band dispersions change largely once including SOC.
The most remarkable difference is that each band without SOC splits into two bands when SOC is considered.
It is the Te vacancy in the monolayer which breaks $\mathcal{I}$ (and $\mathcal{TI}$) inducing the visible Rashba band splitting.
From the orbital-resolved band structures without SOC in Fig.~\ref{fig:1}(c), we find there exist visible band hybridizations between Te-$p_{xy}$ and Te-$p_z$ orbitals around $E_F$.
Using \webirvsp~\cite{GAO2021107760}, the irreducible representations of the high-symmetry \textbf{k} points are calculated and labeled in Fig.~\ref{fig:2}(a). Accordingly, the BR analyses indicate that the two conduction bands belong to $E@1b$ BR, while the highest valence band belongs to $A1@1b$ BR, suggesting the unconventional nature/obstructed atomic limit~\cite{PhysRevB.103.205133,PhysRevResearch.3.L012028,GAO2022598,XYF2021-1}.

SOC often plays important roles to engineering topological states, such as quantum spin Hall effect in graphene~\cite{KM-2005,QSH-graphene} and Ta$_2$M$_3$Te$_5$ (M=Pd,Ni) compounds~\cite{Guo2021,exp2021}, 3D large-SOC-gap topological insulator in Bi$_2$Se$_3$ and NaCaBi families~\cite{Zhang2009,Shao2021}, and so on.
In terms of the the patterned PtTe$_2$ monolayer, once including SOC, the Te-$p_{z}$ dominated band around $\Gamma$ splits due to the Rashba SOC induced by the Te vacancy, as shown in Fig.~\ref{fig:1}(d).
To get more insights in the nontrivial band topology and Rashba SOC band splitting,
we have explored how the band structure evolves with the increasing strength of SOC (denoted by $\lambda$) gradually in Fig.~\ref{fig:2}.
We notice that the nontrivial band topology for $N_e-4$ occupied bands is due to the SOC (can be infinitesimal) induced band gap at $\Gamma$ without involving band inversion ~\cite{Bradlyn2017,DengPRB2022}.
In addition, the nontrivial topologies for $N_e-2$ and $N_e$ occupied bands are due to a gap closing and reopening process as varing $\lambda$.
Taking $N_e-2$ occupied bands as an example, the critical Weyl band crossing between the $(N_e-2)$th band and the $(N_e-1)$th band appears on the M--K line with $\lambda=0.8$, as highlighted by a red dashed ring in Fig.~\ref{fig:2}(e).
Similarly, the WP between the $N_e$th band and the $(N_e+1)$th band appears on the K--$\Gamma$ line with $\lambda=0.98$, as the right inset shown in Fig.~\ref{fig:2}(f).
However, it becomes topologically trivial for $N_e+2$ occupied bands since there are two nontrivial gap openings around both $\Gamma$ and M.

Since the existence of $R_{3z}$ and $M_{100}$ symmetries, both the two classes of WPs above mentioned appear in sextuplet in the first BZ, as shown in Fig.~\ref{fig:3}(a) and Fig.~\ref{fig:3}(b).
Similar with WPs in 3D Weyl semimetals~\cite{PhysRevB.101.155143,GAO2021667,PhysRevLett.124.076403,Shi2021}, these critical WPs also conform to the codimensional analyses.
This can be deduced as follows.
First, both the M--K ($k_y=\pi$) and K--$\Gamma$ ($k_y=0$) lines are $M_{100}*T$ invariant.
In the two-band \textbf{\emph{k $\cdot$ p}} Hamiltonian depicting the Weyl band crossing,
the combined antiunitary symmetry with $[TM_{100}]^2=1$ will reduce the number of the independent $\sigma$ matrices in the \textbf{\emph{k $\cdot$ p}} Hamiltonian to two.
Second, the $k_x$ value in both the $TM_{100}$ invariant lines and the SOC strength $\lambda$ are two tunable parameters to search a WP.
Thus, the number of the independent $\sigma$ matrices in the \textbf{\emph{k $\cdot$ p}} Hamiltonian equals to the number of the tunable parameters,
which indicates that a WP is stable in the 2D parameter space $\{k_x,\lambda\}$.
In other words, a topological phase transition can happen by tuning $\lambda$ in the $M_{100}*T$ invariant lines.
Through the gap closing and reopening process in the evolution, it becomes topologically nontrivial for $N_e~(N_e-2)$ occupied bands.
As a result, we can expected the existence of the helical edge states of the patterned PtTe$_2$ monolayer. The edge spectra are presented in Fig.~\ref{fig:ss-01}(b,d) [in Section C of the SM (SM ~\ref{supthree})].

\subsection{Rashba SOC at $\Gamma$}
\begin{figure*}[!t]
\centering
\includegraphics[width=17 cm]{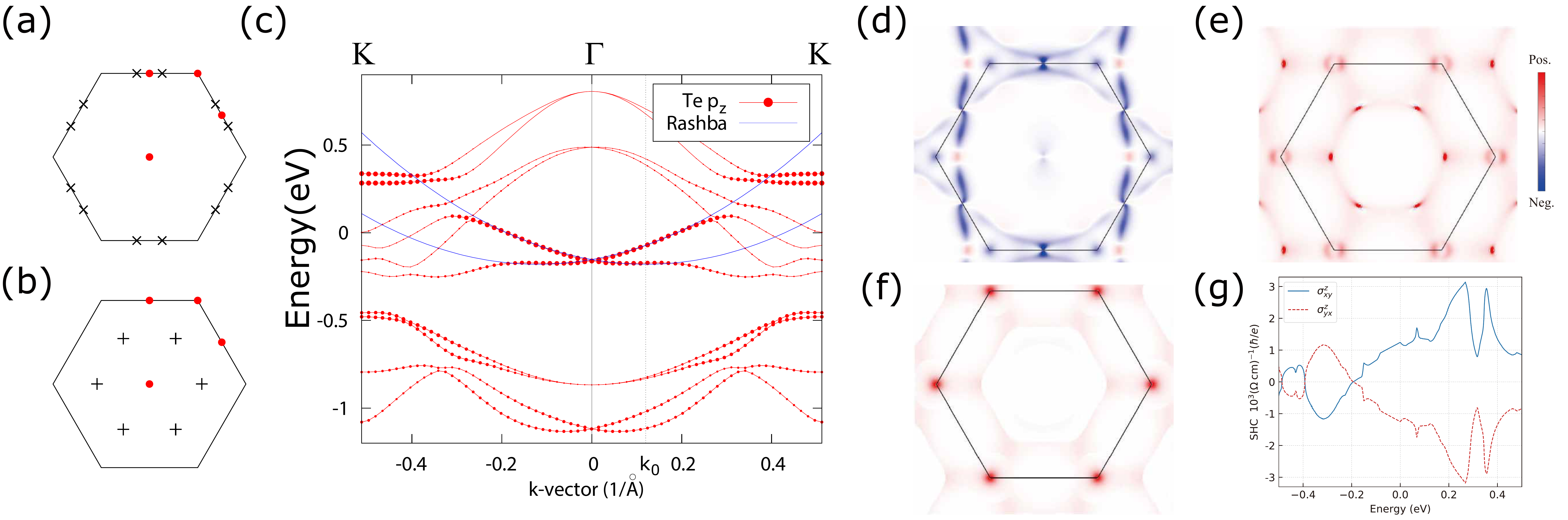}
\caption{(Color online)
The six symmetry-related ($R_{3z}$ and $M_{100}$) WPs formed by band crossings (a) between the $(N_e-2)$th band and the $(N_e-1)$th band with $\lambda = 0.8$ (b) the $N_e$th band and the $(N_e+1)$th band with $\lambda = 0.98$. (c) The Te-$p_z$ projected band structures with SOC, and the two blue lines depicts the parabolically asymptotic behavior of the Rashba SOC induced splitting bands near the $\Gamma$ point.
The distribution of (d) $ \mathcal{O}_{N_{e}-1}(\textbf{\emph{k}})$  and (e) $\mathcal {O}_{N_e}(\textbf{\emph{k}})$ in the 2D BZ.
(f) The distribution of $\sigma_{xy}^{z}$ at $E_F$.
(g) The calculated SHC vs. the chemical potential (ranging from  $E_F-0.5$ eV to  $E_F+0.5$ eV).
}
\label{fig:3}
\end{figure*}

Because the Te vacancy breaks $\mathcal{I}$ in the patterned PtTe$_2$ monolayer,
the Rashba SOC band splitting will appear inevitably.
As the projected band structures with SOC shown in Fig.~\ref{fig:3}(c), the Te-$p_z$ dominated parabolic bands splits clearly near $\Gamma$.
The splitting bands near $\Gamma$ can be well fitted by $E=\frac{[\hbar (k\pm k_0)]^2}{2M}$ with $M=2.02659~m_e$ ($m_e$ denoting the free electronic mass) and $k_0=0.12$ \AA$^{-1}$ (the momentum offset), as the two blue parabolas shown in Fig.~\ref{fig:3}(c).
The coupling strength of the Rashba SOC can be derived as $\alpha_{R}=\frac{2E_{R}}{k_0}=\frac{\hbar^2 k_0}{M}=0.81$ eV\AA.
The estimated $k_0$ is super large in Fig.~\ref{fig:3}(c), as large as the the Bi/Ag(111) surface alloy~\cite{Ag111-Bi-2007}.

\subsection{Large SHC effect}
To explore the intrinsic SHC in the patterned PtTe$_2$ monolayer, the wannier-based TB model under bases of the Te-$p$ and Pt-$d$ orbitals is extracted from the DFT calculations.
As shown in Fig.~\ref{fig:dft-wannier}(a) and Fig.~\ref{fig:dft-wannier}(b), the fitted wannier-based TB bands can reproduce the DFT bands perfectly.
Based on this wannier-based TB model, we have employed the Kubo formula approach at the clean limit~\cite{zhang2017, RevModPhys.82.1959, PhysRevB.47.1651, PhysRevB.48.4442, berry_phase_David} to calculate the SHC of the patterned PtTe$_2$ monolayer,
\begin{equation}
    \begin{aligned}
        \sigma_{\alpha\beta}^{\gamma} = & \frac{e}{\hbar} \sum_n \int_{\text{BZ}} \frac{d\textbf{k}}{(2\pi)^2} f_{n}(\textbf{k}) \Omega^{\gamma}_{\alpha\beta;n}(\textbf{k}), \\
        \Omega^{\gamma}_{\alpha\beta;n}(\textbf{k}) = & 2i\hbar^2 \sum_{m\neq n} \frac{ \langle{u^n_{\textbf{k}}}|{\hat{J}_{\alpha}^{\gamma}}|{u^m_{\textbf{k}}}\rangle\langle{u^m_{\textbf{k}}}|{\hat{v}_{\beta}}|{u^n_{\textbf{k}}}\rangle }{ (\varepsilon^n_{\textbf{k}}-\varepsilon^m_{\textbf{k}})^2 },
    \end{aligned}
\end{equation}
where $\hat{J}^\gamma_\alpha=\frac{1}{2}\{\hat{v}_\alpha,\hat{s}_\gamma\}$ is the spin current operator, with $\hat{s}$ denoting the spin operator, $\hat{v}_{\alpha} = \frac{\partial \hat{H}}{\hbar \partial k_\alpha}$ denoting the velocity operator, and $\alpha, \beta, \gamma=\{x,y,z\}$.
$f_{n}(\textbf{k})$ is the Fermi-Dirac distribution.
$\ket{u_{\textbf{k}}^n}$ and $\varepsilon^n_{\textbf{k}}$ are the eigenvectors and eigenvalues of the TB Hamiltonian respectively.
The distributions of ${\cal O}_N(\textbf{\emph{k}})\equiv \sum_{n=1}^N \Omega^{x}_{yz;n}(\textbf{\emph{k}})$ for $N=N_e -1$ and $N_e$ occupied bands are presented in Fig.~\ref{fig:3}(d) and Fig.~\ref{fig:3}(e), respectively.
As the calculated SHC as a function of the chemical potential shown in Fig.~\ref{fig:3}(g), one can find that the calculated SHC is as large as 1.25$\times 10^3 \frac{\hbar}{e} (\Omega~cm)^{-1}$ at $E=E_F$.
The corresponding distribution at $E=E_F$ is presented in Fig.~\ref{fig:3}(f), which indicates that the large contribution of the SHC at K comes from the SOC band splitting.
In addition, the SHC changes rapidly and monotonically in a wide energy window ranging from $E_F-0.3$ eV to $E_F+0.3$ eV.
At $E-E_F=-0.3 $eV, the SHC changes the sign and becomes $-1.2\times 10^3 \frac{\hbar}{e} (\Omega~cm)^{-1}$,
while at $E-E_F=0.3$ eV, the SHC nearly triples and becomes 3.1$\times 10^3 \frac{\hbar}{e} (\Omega~cm)^{-1}$.
In general, the chemical potential can be tuned by applying gate voltage or introducing chemical doping at the vacancy. 
As shown in Figs.~\ref{fig:dop}(a-c), we can find that the absorption of Tl/Pb at the vacancy behaves as electron dopings, which will increase the E$_F$ with negligible changes in the band structure.
We think our results will benefit to the potential applications in spintronics.

\subsection{Excellent hydrogen evolution reaction activity}
\begin{figure}[!t]
\centering
\includegraphics[width=8.5 cm]{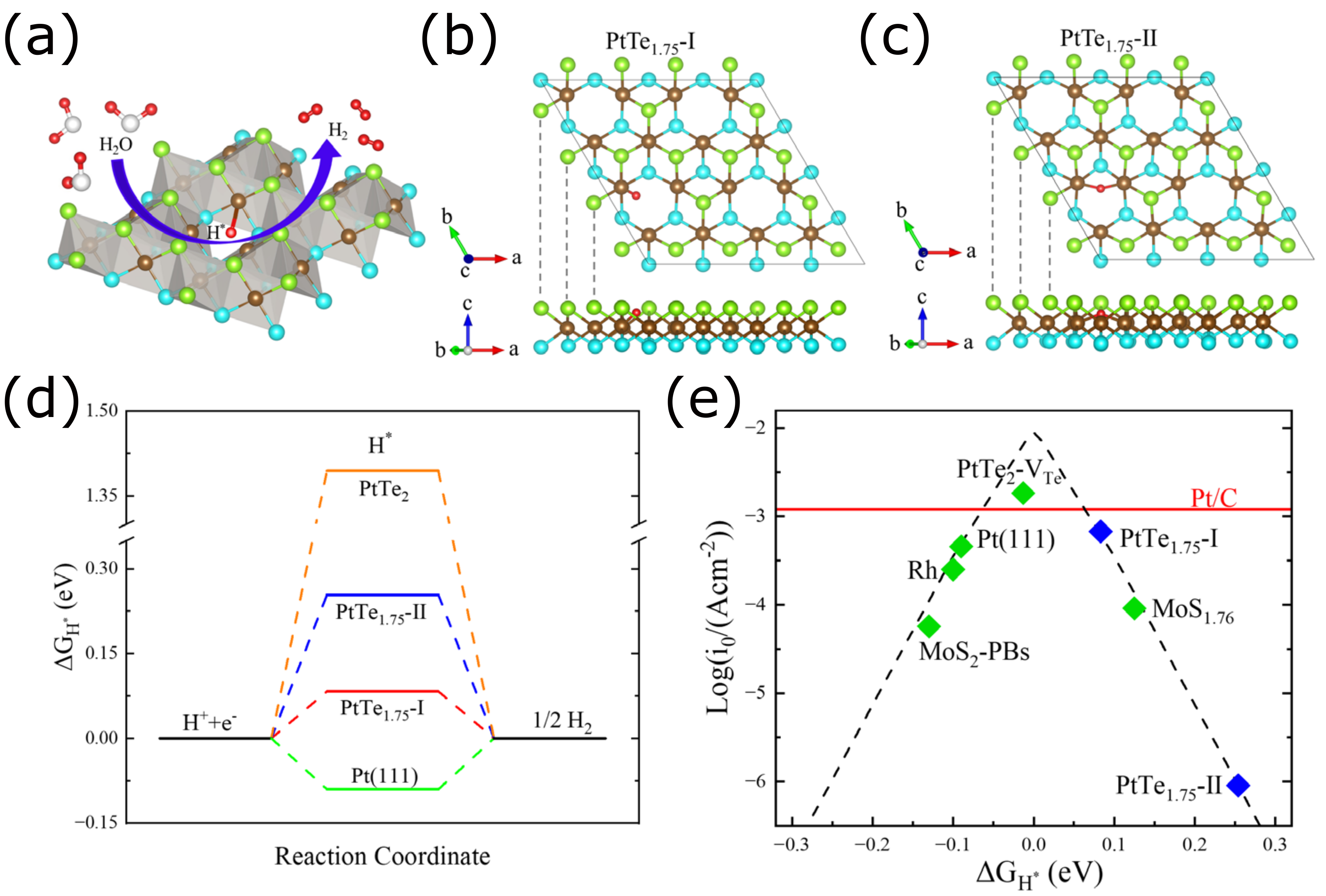}
\caption{(Color online)
(a) Schematic diagram of HER process on PtTe$_{1.75}$ monolayer. Top and side views of (b) the most stable (PtTe$_{1.75}$-I) and (c) metastable (PtTe$_{1.75}$-II) structures after H atom adsorption.
The red ball denotes the absorbed H atom.
(d) Variation of the Gibbs free energy for hydrogen adsorption ($\Delta G_{H^*}$) to different compounds.
(e) Volcano plot depicting the relationship between exchange current density ($i_0$) and $\Delta G_{H^*}$, in which cases of Pt(111)~\cite{N_rskov_2005}, Rh~\cite{N_rskov_2005}, Pt/C~\cite{Yang2021}, PtTe$_2$ with ordered trigonal Te vacancies (PtTe$_{2}$-V$_{Te}$)~\cite{Li2021}, MoS$_{1.76}$~\cite{Jiang2022}, and 2H-1T-phase boundaries of MoS$_2$ (MoS$_2$-PBs)~\cite{Zhu2019} are also included for comparison.
}
\label{fig:4}
\end{figure}

According to the new principle for active catalytic sites~\cite{GAO2022598,XYF2021-2,LGW2021}, the obstructed bulk states in the patterned monolayer (which can be seen as the limit of obstructed surface states) may bring measured catalytic activity.
By exposing undercoordinated atoms as the active sites, vacancy engineering is an important strategy to optimize the HER performance of the basal planes in 2D materials~\cite{Li2021,Jiang2022}. As the acidic HER of the PtTe$_{1.75}$ schematized in Fig.~\ref{fig:4}(a), protons (H$^+$) in solution generate adsorbed H atoms (H$^*$) as intermediate, then the H atoms on the catalyst surface are desorbed to produce hydrogen (H$_2$), which can be formulized as
\begin{equation}
    \begin{aligned}
        H^{+}+e^{-}+*\rightarrow H^{*}\rightarrow \frac{1}{2} H_2+*.
    \end{aligned}
\end{equation}
Here ``*'' denotes some site on the surface, i.e., a ``*" by itself denotes a free site, while H$^{*}$ denotes a hydrogen atom absorbed on the surface. Te vacancy induced states near $E_F$ give PtTe$_{1.75}$ monolayer larger electrical conductivity than pristine PtTe$_2$ monolayer, which will effectively facilitate electron transfer for HER. We used a 2$\times$2 PtTe$_{1.75}$ supercell to simulate the basal plane. Compared with the fully coordinated Te atoms, H atoms are more likely to be adsorbed near the undercoordinated Pt atoms, just as the most stable and metastable structures shown in Fig.~\ref{fig:4}(b) and Fig.~\ref{fig:4}(c). As an important descriptor of HER activity~\cite{N_rskov_2005,Greeley2006}, the change of Gibbs free energy induced by hydrogen adsorption ($\Delta G_{H^*}$) can be defined as~\cite{Norskov2004}

\begin{equation}
    \begin{aligned}
        &\Delta G_{H^*}=G(H^*)-G(*)-\frac{1}{2}G(H_2)\\
        &G=E+ZPE+\int C_{p}  dT -TS,
    \end{aligned}
\end{equation}
where E is internal Energy, ZPE is zero-point energy, $\int C_p  dT$ is the correction of enthalpy, T is temperature, while S denotes entropy. An ideal catalyst for HER should host a near-zero $\Delta G_{H^*}$, which can effectively maintain the balance between adsorption and desorption steps~\cite{Greeley2006}. As shown in Fig.~\ref{fig:4}(d), unlike pristine PtTe${_2}$ monolayer with a large positive $\Delta G_{H^*}$ due to its extremely inert basal plane, the PtTe$_{1.75}$ monolayer hosts an optimal $\Delta G_{H^*}$ (0.08 eV), which is slightly superior to the benchmark material Pt ($\Delta G_{H^*}$ = -0.09 eV)~\cite{N_rskov_2005}. Thus, the active Pt sites induced by Te vacancy greatly optimize hydrogen adsorption in the intermediate, which will significantly improve HER performance~\cite{Prediction2021}. According to N{\o}rskov \emph{et al.}~\cite{N_rskov_2005}, the theoretical exchange current density ($i_0$) as a function of $\Delta G_{H^*}$ is calculated. As shown in Fig.~\ref{fig:4}(e), the PtTe$_{1.75}$ monolayer approaches the volcanic peak from the right with $i_0 = 0.68~mA~cm^{-2}$, which is comparable to commercial Pt/C catalyst ($i_0= 1.2~mA~cm^{-2}$)~\cite{Yang2021}. Therefore, Te vacancy can greatly stimulate the catalytic activity of PtTe$_2$ basal plane and produce excellent HER performance.

\section{Discussion}
The PtTe$_2$ system and its derivatives (the monolayer, multilayer and doped systems) have gained lots of attention since various fascinating properties in them have been proposed, such as type-II Dirac fermion, ultrahigh electrical conductivity, tunability and superconductivity, and so on.
In this work, we find that the PtTe$_{1.75}$ hosts not only unique band structures with three lower-energy bands belonging to $(A1+E)@1b$ BRs (induced by the vacancy), but also large SHC effect and excellent HER performance.
First, we have calculated the corresponding time reversal invariant $\mathbb Z_2$ which indicates the 2D TI nature in the patterned PtTe$_2$ monolayer.
We demonstrate that the topological phase can be deduced by a gap closing and reopening process with the evolution of the strength of SOC from $\lambda =0$ (without SOC) to $\lambda =1.0$ (with SOC).
The critical phase transition occurs at $\lambda =0.98$, which gives a sextuplet of critical WPs.
Second, the Te vacancy breaks $\mathcal{I}$ and induces Rashba SOC band splitting.
The estimated momentum offset is super large with $k_0=0.12$ \AA$^{-1}$.
Third, we find that the SHC is as large as 1.25$\times 10^3 \frac{\hbar}{e} (\Omega~cm)^{-1}$ at $E_F$. 
Furthermore, the SHC varies quickly and almost monotonically from -1.2 to 3.1 $\times 10^3 \frac{\hbar}{e} (\Omega~cm)^{-1}$, indicating that the SHC in the patterned PtTe$_2$ monolayer can be conveniently tuned for various applications.
Lastly, we also find the Te vacancy in the patterned monolayer can induce excellent HER activity.
These results not only offer a new idea to search 2D materials with large SHC, i.e., by introducing inversion-symmetry breaking vacancies in large SOC systems, but also provide a feasible system for the potential application in spintronics and HER catalysts.

\ \\
\noindent \textbf{Acknowledgments}\\
This work was supported by the National Natural Science Foundation of China (Grants No. 11974395, No. 12188101, No. 52188101 and No. 51725103), the Strategic Priority Research Program of Chinese Academy of Sciences (Grant No. XDB33000000), and the Center for Materials Genome.

\ \\
\noindent \textbf{Conflict of Interest}\\
The authors declare no conflict of interest.

\ \\
\noindent \textbf{Data Availability Statement}\\
The datasets used in this article are available from the corresponding author upon request.


%

\clearpage

\begin{widetext}
        \beginsupplement{}
        \setcounter{section}{0}
        \renewcommand{\thesubsection}{\Alph{subsection}}
        \renewcommand{\thesubsubsection}{\alph{subsubsection}}
\section*{SUPPLEMENTARY MATERIAL}

\subsection{Lattice parameters of the patterned PtTe$_2$ monolayer with a Te vacancy}
\label{supone}
The pristine PtTe$_2$ system crystallizes in the CdI$_2$-type trigonal (1T) structure (SG P$\bar{3}$m1) is a layered material stacking along the $z$ axis. Recently, the monolayer structure with kagome lattice formed by one Te vacancy in the 2$\times$2 supercell has been successfully grown~\cite{PtTe2-epitaxial-2021}. The patterned PtTe$_2$ monolayer contains two Te layers, with 4 Te atoms in the bottom layer while 3 Te atoms in the top layer, as shown in Fig.~\ref{fig:2}(a). The corresponding lattice parameters are $a=b=8.1846 {\AA}$, $\alpha=\beta=120^{\circ}$, and the thickness of the vacuum layer along $z$ axis was set to  $30-d_0$ \AA, with $d_0 ~(=2.7253 {\AA})$ denoting the distance between the bottom Te layer and the top Te layer.

\begin{table*}[h!]
    \begin{ruledtabular}
        \caption{%
            Crystal structures of the patterned PtTe$_2$ monolayer in terms of space group P3m1 (SG 156).
            }
        \begin{tabular}{c c c}
            Atoms     & Wyckoff positions  & Fractional coordinates  \\\hline
            Pt1       &     1c             & (0.66667,0.33333,0.0)       \\
            Pt2       &     3d             & (0.16667,0.33333,0.0)       \\
            Te1(bottom)       &     1a             & (0.00000,0.00000,-0.04542)     \\
            Te2(bottom)       &     3d             & (0.50000,0.50000,-0.04542)    \\
            Te3(top)       &     3d             & (0.83333,0.66667,0.04542)     \\
        \end{tabular}
        \label{table:Xnet}
    \end{ruledtabular}
\end{table*}

\subsection{The calculated weak topological invariant $Z_2$}
\label{suptwo}
To characterize the topological properties in the patterned PtTe$_2$ monolayer, the weak topological invariants are calculated by the 1D Wilson loop method. 
Taking Te: 5s$^{2}$5p$^{4}$  and Pt: 5d$^{9}$6s$^{1}$ orbitals into consideration, there are $N_e=82$ valance electrons, resulting in $N_e$ valence bands.
From Figs.~\ref{fig:z2}(a-d), the calculated weak topological invariants for 78, 80, 82 and 84 (corresponds to $N_e -4$, $N_e -2$, $N_e$ and $N_e +2$) occupied bands are $\mathbb Z_{2} = 1$, $\mathbb Z_{2} = 1$, $\mathbb Z_{2} = 1$ and $\mathbb Z_{2} = 0$, respectively.
Thus, the patterned PtTe$_2$ monolayer with $N_e$ valence bands is a 2D TI with $\mathbb Z_{2} = 1$.
\begin{figure}[!htb]
\centering
\includegraphics[width=16.5 cm]{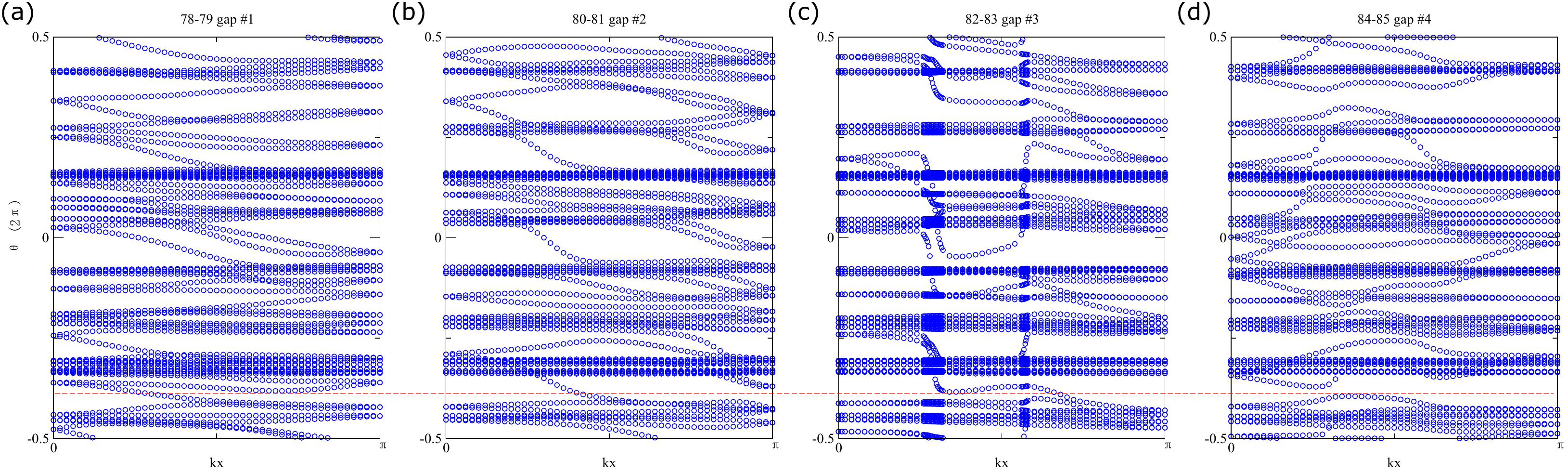}
\caption{(Color online)
The calculated weak topological invariant $\mathbb Z_2$ for (a) 80 (b) 82 (c) 84 and (86) occupied bands, respectively.
}
\label{fig:z2}
\end{figure}

\subsection{Topological surface states of the patterned PtTe$_2$ monolayer}
\label{supthree}
Exotic topological surface states serve as significant fingerprints to identify various topological phases. Based on the tight-binding (TB)
model constructed with the maximally localised Wannier functions and surface Green function methods~\cite{Sancho_1985,RevModPhys.84.1419,WT-Wu2018}, we have
calculated the corresponding surface states to identify the nature of 2D TI.

\begin{figure}[!ht]
\centering
\includegraphics[width=16.5 cm]{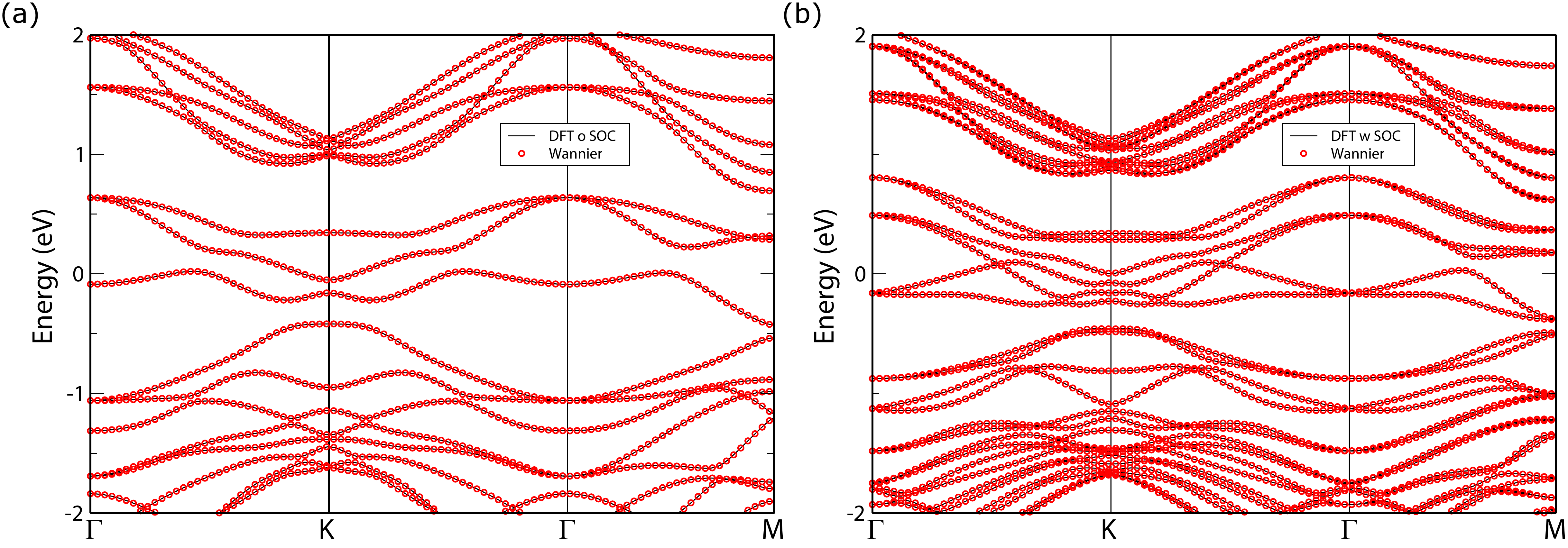}
\caption{(Color online)
DFT vs. Wannier bands (a) without SOC and (b) with SOC.
}
\label{fig:dft-wannier}
\end{figure}

As shown in Fig.\ref{fig:dft-wannier}(a,b), we have chosen Pt-d and Te-p orbitals as the projected bases of the wannier-based TB model, which can reproduce the DFT band structures perfectly.
As expected, the helical edge states corresponds to the 2D TI can be found in the (01) edge states, as shown in Fig.~\ref{fig:ss-01}(b) and Fig.~\ref{fig:ss-01}(d).

\begin{figure}[!hb]
\centering
\includegraphics[width=16.5 cm]{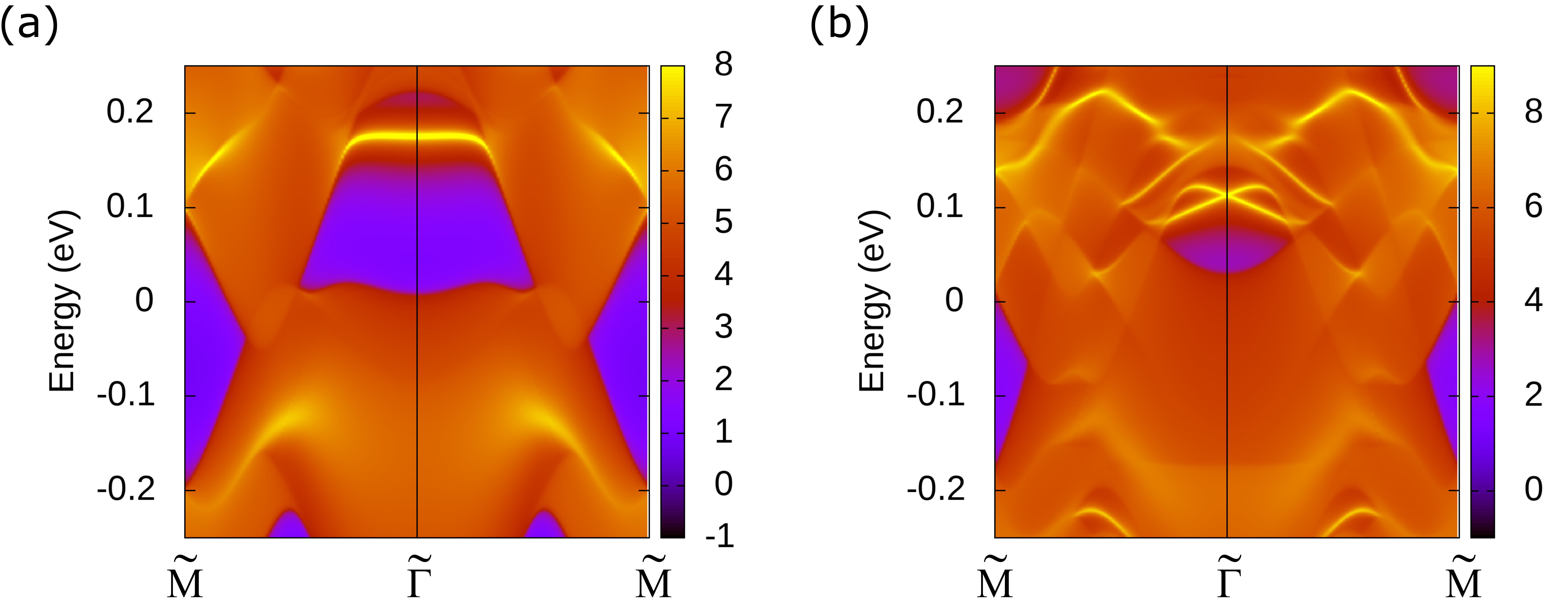}
\includegraphics[width=16.5 cm]{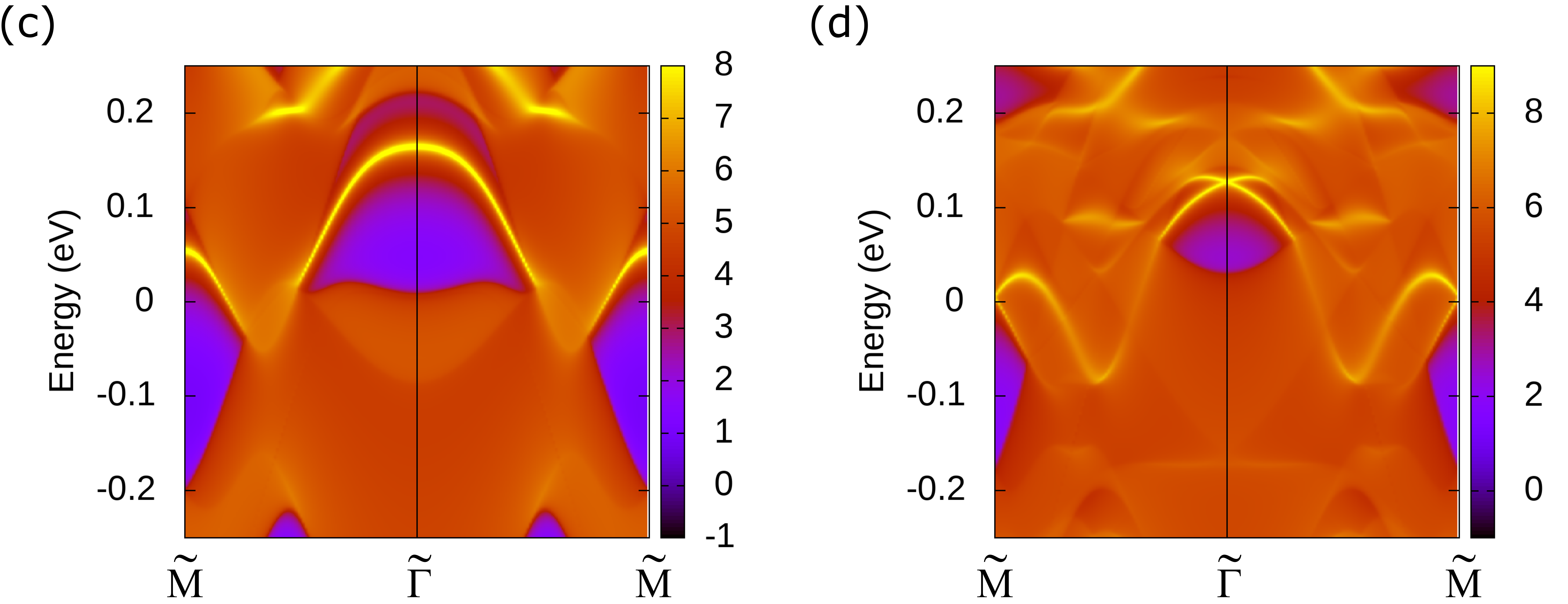}
\caption{(Color online)
The projected edge states along (01) direction (a) without SOC and (b) with SOC in the upper edge.
The projected edge states along (01) direction (c) without SOC and (d) with SOC in the lower edge.
}
\label{fig:ss-01}
\end{figure}

\subsection{Band structures vs. dopping}
Since the electronic properties and SHC of the patterned PtTe$_2$ monolayer are sensitive to the $E_F$, we can adopt different dopings at the Te-vacancy position to tune E$_F$ effectively.

As shown in Figs.~\ref{fig:dop}(a-c), we can find that both the introduced Pb doping and Tl doping behave as electron dopings, which increase the $E_F$ significantly.
The decreased magnitude of the E$_F$ are estimated to be 0.482232 eV and 0.256149 eV, respectively.

\begin{figure}[!hb]
\centering
\includegraphics[width=18 cm]{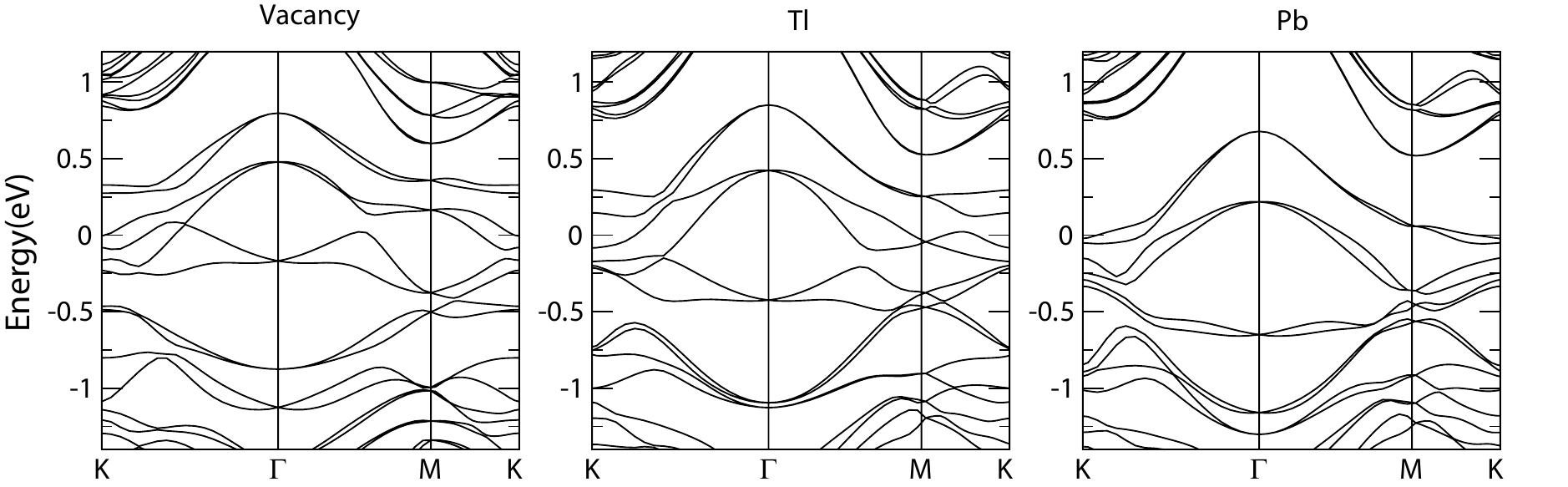}
\caption{(Color online)
Band structures of the patterned PtTe$_2$ monolayer with (a) no doping (Te vacancy), (b) Tl doping, and (c) Pb doping at the Te-vacancy position.
}
\label{fig:dop}
\end{figure}

\label{supfour}
\clearpage

\end{widetext}
\end{document}